\documentclass[aps,prc,preprint,superscriptaddress]{revtex4}

\usepackage{graphicx}
\usepackage{textcomp}
\usepackage{amsfonts}
\usepackage{amssymb}
\usepackage{bm}
\usepackage[dvips]{color}

\begin{document}

\title{Orbital angular momentum of the proton and intrinsic five--quark Fock states}

\author{C. S. An}
\email[]{ancs@swu.edu.cn}
\affiliation{School of Physical Science and Technology, Southwest University, Chongqing 400715, China}

\author{B. Saghai}
\email[]{bsaghai@yahoo.fr}
\affiliation{Institut de Recherche sur les lois Fondamentales de l'Univers, DRF/Irfu,
CEA/Saclay, F-91191 Gif-sur-Yvette, France}

\date{\today}

\begin{abstract}

The orbital angular momentum ($L_q$) of the proton is studied by employing the extended constituent quark model.
Contributions from different flavors, namely, up, down, strange, and charm quarks in the proton are investigated.
Probabilities of the intrinsic $q\bar{q}$  pairs are calculated using a $^{3}P_{0}$ transition operator
to fit the sea flavor asymmetry $I_a=\bar{d}-\bar{u}=0.118\pm0.012$ of the proton~\cite{Towell:2001nh}.
Our numerical results lead to  $L_q=0.158 \pm 0.014$, in agreement with $4/3I_a=0.157 \pm 0.016$,
and consistent with findings based on various other approaches.

\end{abstract}

\maketitle

\section{Introduction}
\label{intro}
%
%
In the late 1980's, the European Muon Collaboration (EMC) published experimental results~\cite{Ashman:1987hv} on the spin asymmetry in
polarized deep inelastic scattering, providing unexpected evidence that the sum of the spins of the quarks add up only to a fraction
of the proton's total spin.
That finding being in contrast to the Gell-Mann--Zweig quark model~\cite{GellMann:1964nj}, in which the spin of the proton is totally
generated by the spins of the three valence quarks, gave rise to the proton spin "crisis".

Since then, efforts aiming at uncovering the spin structure "puzzle" of the nucleon have triggered a significant number of measurements using
various facilities.

In order to emphasize the context of the present work, we start with Ji's sum  rule~\cite{Ji:1996ek}, according to which the nucleon spin
can be decomposed as
\begin{equation}
J_{N}= \sum_{f=u,d,s,c...} ( \small{1/2 \Delta\Sigma_{q}}+ L_q) + J_g\,,
\label{Ji}
\end{equation}
where $1/2 \Delta\Sigma_{q}$ is the contribution from the intrinsic quark spin, $L_q$ the quark orbital angular momentum (OAM) and $J_g$ the
gluon total angular momentum.

In Eq. (\ref{Ji}), the sum over quarks flavors goes beyond the naive constituent quark model (CQM), embodying higher Fock states, namely;
in addition to the conventional nucleon structure with three constituent quarks ($|qqq\rangle$; $q \equiv u,~d$), one introduces
higher Fock five-quark components $|qqqQ\bar{Q}\rangle$, with quark--antiquark pairs $Q \bar{Q} \equiv u \bar{u},~d \bar{d},~s \bar{s},~c \bar{c}...$

The need for the $Q \bar{Q}$ components in the nucleon was emphasized in 1990's by  measurements of the $\bar{d}-\bar{u}$
flavor asymmetry and the ratio ${\bar{u}}/{\bar{d}}$ performed by the New Muon~\cite{Amaudruz:1991at},
E772~\cite{McGaughey:1992kz},  NA51~\cite{Baldit:1994jk}, HERMES~\cite{Ackerstaff:1998sr}, and FNAL E866/NuSea~\cite{Towell:2001nh} Collaborations.
The latest results from each one of the experimental groups using various facilities: BNL, CERN, DESY, Jlab, and SLAC are given in chronological
order in~\cite{Anthony:2000fn}.
In spite of a healthy set of experimental data and intensive theoretical investigations, the question is still open; for recent reviews,
see, e.g., ~\cite{Kuhn:2008sy,Burkardt:2008jw,Aidala:2012mv,Chang:2014jba,Leader:2013jra,Wakamatsu:2014zza,Liu:2015xha,Deur:2018roz}.

Actually, genuine higher Fock states in the baryons' wave functions constitute a pertinent nonperturbative source of the intrinsic quark-antiqurak
components~\cite{Brodsky:1981se}; to be distinguished from the extrinsic pairs arising from gluon splitting
in perturbative QCD and contributing to $J_g$.
The well-known nonvanishing $\bar{d}-\bar{u}$ flavor asymmetry  measured~\cite{Towell:2001nh}, with high enough accuracy, provided stringent
constraints on  the role played by the virtual $Q \bar{Q}$ pairs in the nucleon.
Moreover, while the CQM also predicts a vanishing value for the OAM, the $Q \bar{Q}$ components lead to $L_q  \neq 0$.
Actually, the contribution of the OAM to the spin of proton was found to be comparable to that of the sea quark
( $\approx$ 30\% each)~\cite{Bijker:2011zz}, and much larger than that of the gluons~\cite{Brodsky:2006ha}.

The present work is devoted to studying  the proton's OAM, which continues to be investigated via  various formalisms; see review
papers, e.g., ~\cite{Aidala:2012mv,Leader:2013jra,Wakamatsu:2014zza,Liu:2015xha,Deur:2018roz}.

In phenomenological approaches, based on meson-baryon degrees of freedom, the intrinsic $Q \bar{Q}$ pairs, sea quarks, are handled as a
meson-cloud surrounding the baryon~\cite{Myhrer:2007cf,Huang:2007dy,Garvey:2010fi,Alberg:2012wr}.
Accordingly, the traditional constituent quark model was extended to take into account the Fock components via pionic fluctuations and hence,
generating the measured $\bar{d}-\bar{u}$ flavor asymmetry, and OAM in the nucleon.
The most commonly used configurations embody $N\pi$ and $\Delta\pi$ Fock components in the proton. In this frame, Garvey~\cite{Garvey:2010fi}
obtains $L_q$= 0.147 $\pm$ 0.027.
In~\cite{Nocera:2016zyg}, the relationship between the OAM and the sea flavor asymmetry of the proton in different models was investigated.

Bijker and Santopinto performed a calculation within the unquenched quark model (UQM)~\cite{Bijker:2009up},
based on a quark model with continuum components, to which quark-antiquark pairs are added perturbatively employing a $^{3}P_{0}$
model~\cite{Le Yaouanc:1972ae}. Fixing $J$ = 1/2, they found $L_q$= 0.162.
Lorc{\'e} and Pasquini studied the Wigner distributions in the light cone constituent quark model( LCCQM)~\cite{Lorce:2011kd}, reaching to a
comparable value, $L_q$= 0.126.
Lattice QCD calculations is an ongoing long endeavor; see, e.g.,~\cite{Lin:2017snn,Yang:2019dha}.
Recently, Alexandrou et {\it al.} released the results of a calculation~\cite{Alexandrou:2017oeh} of the quark and gluon
contributions to the proton spin, using an ensemble of gauge configurations with two degenerate light quarks with a mass fixed to approximately
reproduce the physical pion mass.
They found the OAM carried by the quarks in the nucleon to be $L_q$=0.207$\pm$64$\pm$45.
Another recent LQCD calculation by Yang~\cite{Yang:2019dha} lead to smaller central value $L_q$=0.10$\pm$9, but due to the size of the
uncertainties, results from the two investigations turn out to be compatible with each other.

The theoretical frame of the present work is based on an extended chiral constituent quark model (E$\chi$CQM), complemented with the
$SU(6)\otimes O(3)$ symmetry breaking effects.
Recently, the intrinsic sea flavor content including $\bar{u}$, $\bar{d}$, $\bar{s}$ and $\bar{c}$ in the nucleon were investigated
employing our formalism, within which all the possible five-quark Fock components
in the nucleon wave function were taken into account~\cite{An:2012kj,Duan:2016rkr}, coupling between the three- and five-quark
components was assumed to be via $^{3}P_{0}$ quark-antiquark pair creation mechanism~\cite{Le Yaouanc:1972ae}, and the coupling
strength was fixed by fitting~\cite{An:2012kj,Duan:2016rkr} the sea flavor asymmetry of the proton~\cite{Towell:2001nh}.
The corresponding obtained pion-nucleon, strangeness-nucleon~\cite{An:2014aea}, and charm-nucleon sigma terms~\cite{,Duan:2016rkr}
 were found to be reasonably consistent with predictions by the lattice QCD and chiral perturbation theory.

Analogous to the meson-cloud description for the nucleon, the five--quark components in the baryons' wave functions naturally contribute to the
OAM of the proton, required by the angular momentum conservation law.
Consequently, in the present work we study the contributions to the proton' OAM from different quark flavors, by taking into account all
possible five-quark Fock components, based on the results obtained in~\cite{An:2012kj,Duan:2016rkr}.

The present manuscript is organized in the following way: in Sec.~\ref{theo}, we present our theoretical
formalism which includes the wave functions and couplings between three- and five-quark components,
and extract the contributions to the  proton's OAM from relevant five-quark configurations.
We report on our numerical results in Sec.~\ref{num}, and proceed to comparisons with the outcomes of the other approaches
briefly presented above.
Section~\ref{sumcon} contains summary and conclusions.

%
%
\section{Theoretical Frame}
\label{theo}

As shown in~~\cite{An:2012kj,Duan:2016rkr}, considering possible pentaquark components, the
wave function of the proton can be expressed as follows:
\begin{equation}
 |\psi\rangle_{p}=\frac{1}{\mathcal{\sqrt{N}}}\left(|uud\rangle+
 \sum_{i,n_{r},l}C_{in_{r}l}|uud(q\bar{q}),i,n_{r},l\rangle \right)\,,
\label{wfn}
\end{equation}
where the first term is the conventional wave function for the proton with three constituent quarks, and the second one, a sum over all possible
higher Fock components with $q\bar{q}$ pairs, namely, the light, strange, and charm quark-antiquark pairs.
Different possible orbital-flavor-spin-color configurations of the four-quark subsystems in the five-quark system are numbered by $i$; $n_{r}$ and $l$
denote the inner radial and orbital quantum numbers, respectively, as discussed in~\cite{An:2012kj}, the orbital quantum number $l$ in the present
case can only be $1$, and contributions from the configurations with $n_{r}\geq1$ should be negligible, if one takes the coupling between
three- and five-quark components to be via the $^{3}P_{0}$ mechanism, within which the transition operator can be written as
\begin{equation}
 \hat{T}=-\gamma\sum_{j}\mathcal{F}_{j,5}^{00}\mathcal{C}_{j,5}^{00}C_{OFSC}\sum_{m}
\langle1,m;1,-m|00\rangle\chi^{1,m}_{j,5}
\mathcal{Y}^{1,-m}_{j,5}(\vec{p}_{j}-\vec{p}_{5})b^{\dag}(\vec{p}_{j})d^{\dag}(\vec{p}_{5})\,.
\label{op}
\end{equation}
In the above equation, $ \hat{T}$ has units of energy, so that  $\gamma$ is (in natural units)  a dimensionless constant of the model.
$\mathcal{F}_{i,5}^{00}$ and $\mathcal{C}_{i,5}^{00}$ are the flavor and color singlet of the quark-antiquark pair $Q_{i} \bar{Q}_{i} $
in the five-quark system, and $C_{OFSC}$ is an operator to calculate the orbital-flavor-spin-color overlap between the residual
three-quark configuration in the five-quark system and the valence three-quark system.
$\chi^{1,m}_{j,5}$  is a spin triplet wave function with spin $S$=1 and  $ \mathcal{Y}^{1,-m}_{j,5}$ is a solid spherical harmonics
referring to the quark and antiquark  in a relative $P-$wave.
$b^{\dag}(\vec{p}_{j})$ and $d^{\dag}(\vec{p}_{5})$ are the creation operators for a quark and antiquark with momenta
$\vec{p}_{j}$ and $\vec{p}_{5}$, respectively.
The operator $ \hat{T}$,  expressed in second-quantization form, can then be applied in the Fock space.
The coefficient $C_{in_{r}l}$ for a given five--quark component can be related to the transition matrix element  between the  three-
and five-quark configurations of the studied baryon,
\begin{equation}
 C_{in_{r}l}=\frac{\langle uud(q\bar{q}),i,n_{r},l|\hat{T}|uud\rangle}{M_{p}-E_{in_{r}l}}\,,
\end{equation}
where $M_{p}$ is the physical mass of the proton, and $E_{in_{r}l}$ the energy for a
corresponding five-quark component.
In order to estimate the energy splitting for different pentaquark configurations, we employ the chiral constituent quark model
in which the hyperfine interaction between quarks takes the following form:
\begin{eqnarray}
 H_{h}&=&-\sum_{i<j}\vec{\sigma}_{i}\cdot\vec{\sigma}_{j}
                    \Big [ \sum_{a=1}^{3}V_{\pi}(r_{ij})\lambda^{a}_{i}\lambda^{a}_{j}+
                   \sum_{a=4}^{7}V_{K}(r_{ij})\lambda^{a}_{i}\lambda^{a}_{j}+
                   V_{\eta}(r_{ij})\lambda^{8}_{i}\lambda^{8}_{j}
                  +\sum_{a=9}^{12}V_{D}(r_{ij})\lambda^{a}_{i}\lambda^{a}_{j}\nonumber\\
                 && +\sum_{a=13}^{14}V_{D_{s}}(r_{ij})\lambda^{a}_{i}\lambda^{a}_{j}
                  +V_{\eta_{c}}(r_{ij})\lambda^{15}_{i}\lambda^{15}_{j}
                   \Big ]\,,
\label{hyp}
\end{eqnarray}
where $\lambda^{a}_{i}$ denotes the $SU(4)$ Gell-Mann matrix acting on the $i^{th}$ quark, $V_{M}(r_{ij})$ is the potential
of the $M$ meson-exchange interaction between the $i^{th}$ and $j^{th}$ quark, as extensively discussed
in~\cite{Glozman:1995fu}.

Accordingly, there are $17$ different pentaquark configurations (Table~\ref{caco}) forming the Fock
components in the proton wave function. Those 17 configurations are classified into four different categories
according to the orbital and spin symmetry of the four-quark subsystem.
As shown in Table~\ref{caco}, the orbital symmetry for the four-quark subsystem of five-quark components in
the proton can be either the mixed symmetric $[31]_{X}$ or
the completely symmetric $[4]_{X}$; the general wave functions for these two different kinds of
pentaquark configurations with a spin projection $+1/2$ can be written as~\cite{An:2005cj}
\begin{eqnarray}
|uud(q\bar{q}),i,0,1;+1/2\rangle &=&
\sum_{abcde}\sum_{Ms'_{z}ms_{z}}C^{\frac{1}{2}\frac{1}{2}}
_{JM,\frac{1}{2}s'_{z}}C^{JM}_{1m,Ss_{z}}C^{[1^{4}]}_{[31]_{a}
[211]_{a}}C^{[31]_{a}}_{[31]_{b}[FS]_{c}}
C^{[FS_{i}]_{c}}_{[F_{i}]_{d}[S_{i}]_{e}}
[31]_{X,m}(b)
\nonumber\\
&&[F_{i}]_{d}[S_{i}]_{s_{z}}(e)[211]_{C}(a)\bar\chi_{s'_{z}}
\varphi(\{\vec{r}_q\}),~i=1,\cdots,10\,,\\
|uud(q\bar{q}),i,0,1;+1/2\rangle &=&
\sum_{abc}\sum_{s_{z}mm's'_{z}}C^{\frac{1}{2}
\frac{1}{2}}_{1s_{z},jm}C^{jm}_{1m',\frac{1}{2}s'_{z}}C^{[1^4]}_{[31]_{a}
[211]_{a}}C^{[31]_{a}}_{[F_{i}]_{b}[S_{i}]_{c}}[F_{i}]_{b}[S_{i}]_{c}[211]_{C,a}\nonumber\\
&&\bar Y_{1m'} \bar \chi_{s_{z'}}\varphi(\{\vec r_q\})~i=11,\cdots,17\,,
\label{wfc}
\end{eqnarray}
respectively. Here, [F ], [S], and [211] correspond to the flavor, spin and color state wave functions, denoted by their relevant Weyl tableaux;
$\bar Y_{1m'}$ and $ \bar \chi_{s_{z'}}$ refer to the orbital and spin states, respectively.
%
%
\begin{table*}[t]
\caption{\footnotesize Categories (2nd line) and associated configurations (lines 3-8) for five-quark components.
\label{caco}}
%
\begin{tabular}{ccccccccccc}
\hline\hline
i & Category / Config. && i & Category / Config. && i & Category / Config. && i & Category / Config.  \\
  & I / $[31]_{X}[22]_{S}$ &&  & II / $[31]_{X}[31]_{S}$ &&  & III / $[4]_{X}[22]_{S}$ &&  & IV / $[4]_{X}[31]_{S}$  \\
\hline
 1 & $[31]_{X}[4]_{FS}[22]_{F}[22]_{S}$    &&  5 & $[31]_{X}[4]_{FS}[31]^1_{F}[31]_{S}$ &&
 11 & $[4]_{X}[31]_{FS}[211]_{F}[22]_{S}$ &&  14 &$[4]_{X}[31]_{FS}[211]_{F}[31]_{S}$ \\
 2 & $[31]_{X}[31]_{FS}[211]_{F}[22]_{S}$  && 6 & $[31]_{X}[4]_{FS}[31]^2_{F}[31]_{S}$ &&
 12 & $[4]_{X}[31]_{FS}[31]^1_{F}[22]_{S}$&&  15 & $[4]_{X}[31]_{FS}[22]_{F}[31]_{S}$ \\
 3 & $[31]_{X}[31]_{FS}[31]^1_{F}[22]_{S}$ && 7 & $[31]_{X}[31]_{FS}[211]_{F}[31]_{S}$ &&
 13 & $[4]_{X}[31]_{FS}[31]^2_{F}[22]_{S}$ &&  16 & $[4]_{X}[31]_{FS}[31]^1_{F}[31]_{S}$ \\
 4 & $[31]_{X}[31]_{FS}[31]^2_{F}[22]_{S}$ && 8 & $[31]_{X}[31]_{FS}[22]_{F}[31]_{S}$ && & &&  17 & $[4]_{X}[31]_{FS}[31]^2_{F}[31]_{S}$ \\
   &                                       && 9 & $[31]_{X}[31]_{FS}[31]^1_{F}[31]_{S}$ && & && & \\
   &                                       && 10 & $[31]_{X}[31]_{FS}[31]^2_{F}[31]_{S}$ && & && & \\
\hline
\hline
\end{tabular}
\end{table*}
Considering the flavor symmetry of the four-quark subsystem, $[31]_{F}^{1}$ limits the
quark-antiquark pair in the pentaquark configurations to be $u\bar{u}$ or $d\bar{d}$,
while $[31]_{F}^{2}$ and $[211]_{F}$ rule out the pentaquark configurations with a light quark-antiquark
pair.

At this point, we discuss the OAM possibly arising from each of the four categories in Table~\ref{caco}.
In category I, the spin symmetry of the four-quark subsystem is $[22]_{S}$, which leads to the spin quantum number
$S=0$.
It is straightforward to show that the projections of the quark orbital angular momentum arising from all the four configurations
are the same,
\begin{equation}
\langle uud(q\bar{q}),i,0,1;+1/2|\hat{L}_{qz}|uud(q\bar{q}),i,0,1;+1/2\rangle=2/3C_{in_{r}l}^{2}/\mathcal{N},~i=1,\cdots,4\,.
\end{equation}
Note that we have taken the notation,
\begin{equation}
\hat{L}_{qz}=\sum_{f}\hat{l}_{f+\bar{f}}=\sum_{f}(\hat{l}_{f}+\hat{l}_{\bar{f}})_{z}\,,
\end{equation}
where $\hat{l}_{f}$ and $\hat{l}_{\bar{f}}$ are the OAM operators for the quark and antiquark with a flavor $f$, respectively, and
the sum runs over the flavors $u$, $d$, $s$, and $c$.

The four configurations in category I contribute differently to the proton sea flavor asymmetry.
Taking the flavor $SU(3)$ symmetry for light and strange quarks, and neglecting the five-quark components with
a $c\bar{c}$ pair in the proton, then respective contributions to $I_a=\bar{d}-\bar{u}$ due to the four configurations in category I read
\begin{equation}
I_{a,1}=2/3C_{1n_{r}l}^{2}/\mathcal{N},~I_{a,2}=0,~I_{a,3}=-1/3C_{3n_{r}l}^{2}/\mathcal{N},~I_{a,4}=0\,.
\end{equation}
Here, we have labeled the contribution from the $i$th five-quark configuration as $I_{a,i}$, and hereafter, we will take the same
convention for the other configurations.

In category II, the spin symmetry of the four-quark subsystem is $[31]_{S}$, which leads to the spin quantum number $S=1$.
Coupling between spin $S=1$ and orbital angular momentum $L=1$ of the four-quark subsystem leads to the total angular momentum
$J_{4}$ equal to  $0$ or $1$.
In the present work, we take $J_{4}=0$  because of the lower energy.
Then, one finds that the  projections of the quark orbital angular momentum arising from all the configurations in category II vanish
\begin{equation}
\langle uud(q\bar{q}),i,0,1;+1/2|\hat{L}_{qz}|uud(q\bar{q}),i,0,1;+1/2\rangle=0,~i=5,\cdots,10\,.
\end{equation}

For the six configurations in category II, respective contributions to the proton sea flavor asymmetry are,
\begin{equation}
I_{a,5}=-1/3C_{5n_{r}l}^{2}/\mathcal{N},~I_{a,6}=I_{a,7}=I_{a,10}=0,~I_{a,8}=2/3C_{8n_{r}l}^{2}/\mathcal{N},~I_{a,9}=-1/3C_{9n_{r}l}^{2}/\mathcal{N}\,,
\end{equation}
according to the flavor structure of the corresponding configuration.

In categories III and IV, the orbital wave function for the four-quark subsystem is $[4]_{X}$, namely, the orbital angular momentum of the
four-quark subsystem is $L=0$.
And the antiquark is in its first orbitally excited state in the present case.
Therefore, contributions to the proton angular momentum by configurations in categories III and IV should be from the antiquark,
\begin{eqnarray}
\langle uud(q\bar{q}),i,0,1;+1/2|\hat{L}_{qz}|uud(q\bar{q}),i,0,1;+1/2\rangle &=& 2/3C_{in_{r}l}^{2}/\mathcal{N},~i=11,\cdots,13\,,\\
\langle uud(q\bar{q}),i,0,1;+1/2|\hat{L}_{qz}|uud(q\bar{q}),i,0,1;+1/2\rangle &=& 0,~i=14,\cdots,17\,.
\end{eqnarray}
Moreover, it is straightforward to show that
\begin{eqnarray}
& I_{a,11}=0,~I_{a,12}=-1/3C_{12n_{r}l}^{2}/\mathcal{N},~I_{a,13}=0\,,\\
& I_{a,14}=0,~I_{a,15}=2/3C_{15n_{r}l}^{2}/\mathcal{N},~I_{a,16}=-1/3C_{16n_{r}l}^{2}/\mathcal{N},~I_{a,17}=0\,.
\end{eqnarray}

Accordingly, the projection of the proton OAM reads
\begin{equation}
_{p}\langle\psi;+1/2|\hat{L}_{qz}|\psi;+1/2\rangle_{p}=\frac{2}{3\mathcal{N}}\left(\sum_{i=1,4}C_{in_{r}l}^{2}+\sum_{i=11,13}C_{in_{r}l}^{2}
\right) \,,
\label{plz}
\end{equation}
and the flavor asymmetry of the proton takes the following form:
\begin{equation}
I_{a}=\bar{d}-\bar{u}=\frac{2}{3\mathcal{N}}\left(C_{1n_{r}l}^{2}+C_{8n_{r}l}^{2}+C_{15n_{r}l}^{2}\right)-
\frac{1}{3\mathcal{N}}\left(C_{3n_{r}l}^{2}+C_{5n_{r}l}^{2}+C_{9n_{r}l}^{2}+C_{12n_{r}l}^{2}+C_{16n_{r}l}^{2}\right)\,.
\label{fas}
\end{equation}
It is obvious that in the present approach, projections of the OAM and flavor asymmetry of the proton are not equivalent
to each other.
Finally, one has to note that the flavor asymmetry $I_{a}$ given in (\ref{fas}) is obtained by neglecting the five-quark components
with a charm quark-antiquark pair and taking $SU(3)$ flavor symmetry for light and strange quarks.
In any case, since probabilities for the five-quark components with strange and charm quark-antiquark pairs
in the nucleon should not be significantly large, one can expect that projection of the OAM should be slightly larger than
the flavor asymmetry according to Eqs.~(\ref{plz}) and (\ref{fas}).

%
%
\section{Numerical results and discussion}
\label{num}

To get the numerical results, one has to determine the probabilities for all the light, strangeness and charmness
components in the proton, as discussed in Refs.~\cite{An:2012kj,Duan:2016rkr,An:2014aea}. They depend on the coupling
strengths $V$ for Goldstone boson exchanges, the degenerated energy $E_{0}$ for different pentaquark configurations,
when differences between the light, strange and charm quark constituent masses, flavor $SU(4)$ symmetry breaking effects and
hyperfine interactions between quarks are not included, and the general orbital overlap factor
$V\propto\langle uud(q\bar{q}),i,n_{r},l|\hat{T}|uud\rangle$.
Same as in~\cite{An:2012kj}, here the parameters $V$ for Goldstone boson exchange model are taken to be the empirical
values~\cite{Glozman:1995fu}.
$E_{0}=2127$~MeV is also an empirical value~\cite{An:2012kj}, and $V$ was determined by fitting~\cite{An:2012kj,Duan:2016rkr}
the sea flavor asymmetry of the proton  $I_{a}^{exp}=0.118\pm0.012$~\cite{Towell:2001nh}, resulting in
\begin{equation}
V=572\pm47~MeV\,.
\label{V}
\end{equation}

With the parameters given above, one  obtains the probabilities for the five-quark Fock components in
the proton wave function; the numerical values were reported in~\cite{Duan:2016rkr}.
%
\begin{table*}[b]
\caption{\footnotesize Contributions to the projection of the proton's OAM from different flavors for quark and antiquark components
$\langle \hat{l}_{f+\bar{f}}\rangle_{q}^{i}$, per five-quark Fock configuration $i$, with probability $\mathcal{P}_{q}^{i}$ (in \%).
\label{nr}}
\renewcommand
\tabcolsep{0.08cm}
\renewcommand{\arraystretch}{1.2}
\scriptsize
\begin{tabular}{cccccccccccccccc}
\hline\hline
       &                              \multicolumn{5}{c}{light}   &\multicolumn{5}{c}{strangeness} &\multicolumn{5}{c}{charmness}             \\

\hline

  i & $\mathcal{P}_{l}^{i}$ & $\langle \hat{l}_{u+\bar{u}}\rangle_{l}^{i}$  & $\langle\hat{l}_{d+\bar{d}}\rangle_{l}^{i}$  & $\langle \hat{l}_{s+\bar{s}}\rangle_{l}^{i}$ &$\langle\hat{l}_{c+\bar{c}}\rangle_{l}^{i}$
              & $\mathcal{P}_{s}^{i}$ & $\langle \hat{l}_{u+\bar{u}}\rangle_{s}^{i}$  & $\langle\hat{l}_{d+\bar{d}}\rangle_{s}^{i}$  & $\langle \hat{l}_{s+\bar{s}}\rangle_{s}^{i}$ &$\langle\hat{l}_{c+\bar{c}}\rangle_{s}^{i}$
              & $\mathcal{P}_{c}^{i}$ & $\langle \hat{l}_{u+\bar{u}}\rangle_{c}^{i}$  & $\langle\hat{l}_{d+\bar{d}}\rangle_{c}^{i}$  & $\langle \hat{l}_{s+\bar{s}}\rangle_{c}^{i}$ &$\langle\hat{l}_{c+\bar{c}}\rangle_{c}^{i}$\\

\hline

     1 & 14.62 (1.21)& $0.333$  & $0.333$  & $0$  & $0$
        &  0.98 (8)      & $0.333$  & $0.167$  & $0.167$  & $0$
        & 0.04 (0)      & $0.333$  & $0.167$  & $0$  & $0.167$\\

     2 & 0 & $0$  & $0$  & $0$  & $0$
        & 0.36 (3)       & $0.250$  & $0.208$  & $0.208$  & $0$
        & 0.03 (1)      & $0.250$  & $0.208$  & $0$  & $0.208$\\

     3 & 1.65 (14)  & $0.500$  & $0.167$  & $0$  & $0$
        & 0      & $0$ & $0$  & $0$ & $0$
        & 0      & $0$  & $0$  & $0$  & $0$ \\

     4 & 0  &  $0$  & $0$  & $0$  & $0$
        & 0.26 (2)      & $0.361$  & $0.097$  & $0.208$  & $0$
        & 0.03 (1)      & $0.361$  & $0.097$   & $0$  &  $0.208$\\

        \\

     11 & 0& 0  & 0  & $0$  & $0$
        &  0.85 (8)      & 0  & 0  & $0.667$  & $0$
        & 0.09 (1)      & 0  & 0  & $0$  & $0.667$\\

     12 & 4.14 (37) & 0.444  & 0.222  & $0$  & $0$
        & 0      & 0  & 0  & 0  & 0
        & 0     & 0  & 0  & 0  & 0\\

     13 & 0  & 0  & 0  & $0$  & $0$
        & 0.65 (6)      & $0$ & $0$  & $0.667$ & $0$
        & 0.09(0)      & $0$  & $0$  & $0$  & $0.667$ \\

\hline
\hline
\end{tabular}
\end{table*}

As discussed in Sec.~\ref{theo}, the pentaquark configurations in categories II and IV cannot contribute to the projection of the OAM,
since the total angular momentum $J_{4}=0$ for the four-quark subsystem in category II and  $J_{5}=0$ for the antiquark
in category IV.

In  our previous studies on the strangeness magnetic form factor of the proton~\cite{An:2013daa} and the nonperturbative strangeness
suppression~\cite{An:2017flb}, which successfully reproduced the relevant data, all four categories intervene.
But, in the present case,  only the pentaquark configurations in categories I and III contribute to the OAM.
Accordingly, the expectation values for the projection of the OAM of different flavors reads
\begin{equation}
\langle \hat{l}_{f+\bar{f}}\rangle_{q}^{i}=\langle uud(q\bar{q}),i,0,1;+1/2|\hat{l}_{f+\bar{f}}|uud(q\bar{q}),i,0,1;+1/2\rangle\,,
\end{equation}
with $i=1,\cdots4;11,\cdots13$, $f=u,d,s,c$ denoting contributions from different flavors, and the subscript $q=l,s,c$ denoting
contributions from the light, strangeness and charmness components in the proton (Table~\ref{nr}).
In addition, the corresponding probabilities for the five-quark Fock components are also listed in columns $\mathcal{P}_{l}^{i}$,
$\mathcal{P}_{s}^{i}$ and $\mathcal{P}_{c}^{i}$ in Table~\ref{nr}.
Accordingly, the OAM per flavor reads
\begin{equation}
\langle\hat{l}_{f+\bar{f}}\rangle=\sum_{i=1}^{4}\left(\mathcal{P}_{l}^{i}\langle\hat{l}_{f+\bar{f}}\rangle_{l}^{i}
+\mathcal{P}_{s}^{i}\langle\hat{l}_{f+\bar{f}}\rangle_{s}^{i}
+\mathcal{P}_{c}^{i}\langle\hat{l}_{f+\bar{f}}\rangle_{c}^{i}\right) +
\sum_{i=11}^{13}\left(\mathcal{P}_{l}^{i}\langle\hat{l}_{f+\bar{f}}\rangle_{l}^{i}
+\mathcal{P}_{s}^{i}\langle\hat{l}_{f+\bar{f}}\rangle_{s}^{i}
+\mathcal{P}_{c}^{i}\langle\hat{l}_{f+\bar{f}}\rangle_{c}^{i}\right)\,.
\end{equation}
Accordingly, contributions to the projection of the OAM of the proton from different flavors are as follows:
\begin{eqnarray}
\langle\hat{l}_{u+\bar{u}}\rangle&=&0.081\pm0.007\,,\nonumber\\
\langle\hat{l}_{d+\bar{d}}\rangle&=&0.063\pm0.006\,,\nonumber\\
\langle\hat{l}_{s+\bar{s}}\rangle&=&0.013\pm0.001\,,\nonumber\\
\langle\hat{l}_{c+\bar{c}}\rangle&\simeq&0.001\pm0.000\,.
\end{eqnarray}

Contributions from up and down quarks to the projection of the proton's OAM are roughly in the same range,Table~\ref{nr},
while $\langle\hat{l}_{d+\bar{d}}\rangle$ is slightly smaller, and those from the strange and charm quarks are much smaller.
In total, one gets
\begin{equation}
L_q  \equiv  \ \langle p,+1/2|\hat{L}_{qz}|p,+1/2\rangle=\langle\hat{l}_{u+\bar{u}}\rangle+
\langle\hat{l}_{d+\bar{d}}\rangle
+\langle\hat{l}_{s+\bar{s}}\rangle
+\langle\hat{l}_{s+\bar{s}}\rangle
=0.158\pm0.014\,,
\end{equation}
and then the relation between the orbital angular momentum and the sea flavor asymmetry, as expected, reads
\begin{equation}
\langle p,+1/2|\hat{L}_{qz}|p,+1/2\rangle\simeq4/3 I_{a}\,.
\end{equation}

As briefly presented in the Introduction, the quark contributions to the proton OAM and the spin structure of the nucleon
have been intensively investigated, using different approaches.
In Table~\ref{cdm}, we compare our numerical results to those recently reported within other approaches.

%
\begin{table}[h]
\caption{\footnotesize The OAM of the proton in different models.
\label{cdm}}
\renewcommand
\tabcolsep{0.50cm}
\renewcommand{\arraystretch}{1.2}
\scriptsize
\begin{tabular}{cccccc}
\hline\hline
    Approach [Ref.]        & $\langle\hat{l}_{u+\bar{u}}\rangle$ &  $\langle\hat{l}_{d+\bar{d}}\rangle$  & $\langle\hat{l}_{s+\bar{s}}\rangle$  &  $\langle\hat{l}_{c+\bar{c}}\rangle$  & $L_q \equiv \langle\hat{L}_{qz}\rangle$  \\
\hline

E$\chi$CQM [Present work]      &  0.081(7) &  0.063(5)  &  0.013(2)   &  0.001(0)  &      0.159(14)     \\

UQM~\cite{Bijker:2009up}         &          $-$                  &       $-$                 &      $-$                   &          $-$                 &      0.162              \\

LCCQM~\cite{Lorce:2011kd}   &   0.071            &   0.055            &   $-$              &   $-$       &       0.126  \\

$\pi$-Cloud~\cite{Garvey:2010fi} &  $-$               &  $-$                   &  $-$                      &   $-$                       &      0.147                   \\

LQCD~\cite{Alexandrou:2017oeh}  &   -0.107(40)            &   0.247(38)            &   0.067(21)              &  $-$            &       0.207(78)         \\

LQCD~\cite{Yang:2019dha}   &   -0.14(4)            &   0.20(3)            &   0. 04(2)             &   $-$          &       0.10(9)          \\
\hline
\hline
\end{tabular}
\end{table}

In the naive  quark model, since all the constituent quarks in the proton are in their ground states, the projections
of the OAM due to both up and down quarks are zero.

In~\cite{Bijker:2009up}, the nucleon orbital angular momentum is investigated using the unquenched quark model (UQM),
within which the effects of the quark-antiquark pairs including $u\bar{u}$, $d\bar{d}$ and $s\bar{s}$ are taken
into account, and the quark-antiquark pairs creation is assumed to be via a $^{3}P_{0}$ mechanism.
Their findings show that the quark-antiquark pairs have sizable contributions to the proton OAM.
Their numerical result  and ours are in (almost) perfect agreement, although the contributions per flavor are not given in~\cite{Bijker:2009up}.

Within the meson-cloud picture, as discussed in Sec.~\ref{intro}, if one only considers the $N\pi$ and $\Delta\pi$ Fock components
in the proton, the projection of the OAM of the proton should be equal to the proton flavor asymmetry $\bar{d}-\bar{u}$,
as studied in~\cite{Garvey:2010fi},  i.e. $L_q\sim0.147$, consistent with our result within $1\sigma$.

The quarks contribution to the OAM was also obtained from the Wigner distribution for unpolarized quarks in the longitudinally polarized nucleon.
The formalism is  applied in the light cone constituent quark model (LCCQM), leading  to a compatible value with ours, within $2-3\sigma$.

Numerical values for $L_q$ within the lattice QCD calculations were reported.
Here, a caution is in order: in the present model, contributions to the proton OAM are exclusively due to the intrinsic
sea content $q\bar{q}$, while LQCD approaches embody also extrinsic quark-antiquark pairs arising from the gluon splitting in the perturbative
QCD regime ($g \to q\bar{q}$).
Nevertheless, in Table~\ref{cdm}, we report results from two approaches~\cite{Alexandrou:2017oeh,Yang:2019dha}.
The first remark is that contributions per flavor for light quarks are very different from our values, as well as from those obtained within LCCQM.
For the strangeness components, discrepancies are around $2\sigma$.
However, given the rather large uncertainties in the LQCD results, the sum  over all contributions turns out to be consistent, within
$1\sigma$, with  all other values reported in Table~\ref{cdm}.
Accordingly, a meaningful comparison would require separating in the LQCD calculations contributions from intrinsic and extrinsic
quark-antiquark pairs, and reducing significantly the uncertainties, which is a huge task.

%
%
\section{Summary and Conclusions}
\label{sumcon}

To summarize, in the present work, we investigate the OAM of the proton by taking into account all the possible light, strangeness, and
charmness five-quark Fock components in the wave function of proton.
Coupling between three- and five-quark components was dealt with via the $^{3}P_{0}$ quark-antiquark pairs creation mechanism; the model parameters
are empirical values~\cite{An:2012kj,Glozman:1995fu}.
The only adjusted parameter,  $V$ in Eq. (\ref{V}), for the Goldstone boson exchange model, was determined by
fitting~\cite{An:2012kj,Duan:2016rkr} the experimental data for the sea flavor asymmetry $I_a=\bar{d}-\bar{u}=0.118\pm0.012$ of the
proton~\cite{Towell:2001nh}.
This ensemble allowed us postdicting, on the one hand, the strangeness magnetic moment $\mu_s$ and
the strangeness magnetic moment $G^s_M $ of the proton~\cite{An:2013daa},
and on other hand, shedding a light~\cite{An:2017flb} on the  measured~\cite{Park:2014zra}  quark-antiquark ratios $r_\ell={u\bar{u}} / {d\bar{d}}$,
$r_s={s\bar{s}} / {d\bar{d}}$, and the strangeness content of the proton $\kappa_s=2{s \bar{s}} / ({u\bar{u}} + {d \bar{d}})$.

In the present work, we  studied the complete set of the 17 five-quark configurations, falling in four categories and showed that
only seven configurations in two of the categories contribute to the OAM.
Accordingly, the proton OAM carried by quarks turns out be $L_q=0.158 \pm 0.014$ in our model,  in perfect  agreement with $4/3I_a=0.157 \pm 0.016$,
as expected.
Contributions from the up and down quarks and antiquarks are the dominant ones and comparable to each other, while those from strange and charm
quarks and antiquarks are rather small.

We proceeded to comparisons between our results and recent findings within other approaches.
Perfect agreement was obtained with the result coming from the unquenched quark model~\cite{Bijker:2009up}.
The meson-cloud picture, embodying the $N\pi$ and $\Delta\pi$ Fock components  in the proton, leads to a value~\cite{Garvey:2010fi}
consistent with ours within $1\sigma$. That is also the case with respect to the LQCD~\cite{Alexandrou:2017oeh,Yang:2019dha}, albeit
with rather large uncertainties.
The light cone constituent quark model's outcome is compatible with ours, within $2-3\sigma$.

In conclusion, our determination of the proton's OAM falls reasonably well in the range of values reported by other authors, underlining the
crucial role played by intrinsic five-quark components in the proton.

%
%
\begin{acknowledgments}

This work is partly supported by the National Natural Science Foundation of China under Grant No. 11675131.

\end{acknowledgments}

\newpage

%

%
%
%
\end{document}